\documentclass[a4paper,12pt,tightenlines,notitlepage]{revtex4-1}
\usepackage{amsmath,amssymb}
\usepackage{color}
\usepackage{graphicx}
\definecolor{darkred}{rgb}{0.7,0,0}
\definecolor{darkgreen}{rgb}{0,0.3,0}
\definecolor{darkblue}{rgb}{0,0,0.7}
\definecolor{darkmagenta}{rgb}{0.5,0,0.5}
\usepackage[pdftex,unicode,colorlinks,citecolor=darkblue,linkcolor=darkblue,bookmarks=true]{hyperref}

\newcommand{\bra}[1]{\langle#1|}
\newcommand{\ket}[1]{|#1\rangle}

\newcommand{\mean}[1]{\langle#1\rangle}
\newcommand{\Tr}{\mathop{\rm Tr}\nolimits}
\newcommand{\svector}[2]{\begin{pmatrix}#1 \\ #2 \end{pmatrix}}
\newcommand{\smatrix}[4]{\begin{pmatrix}#1 & #2 \\ #3 & #4\end{pmatrix}}
\newcommand{\partd}[2]{\dfrac{\partial#1}{\partial#2}}
\newcommand{\intinfty}{\displaystyle\int_{-\infty}^{\infty}\!}
\newcommand{\intOinfty}{\displaystyle\int_{0}^{\infty}\!}
\newcommand{\sinc}{\mathop{\rm sinc}\nolimits}
\renewcommand{\Re}{\mathop{\rm Re}\nolimits}
\renewcommand{\Im}{\mathop{\rm Im}\nolimits}

\begin{document}

\title{Non-classical features of Polarization Quasi-Probability Distribution}

\author{M.V.Chekhova}
\affiliation{Max Planck Institute for the Science of Light, G\"unther-Scharowsky-Stra\ss{}e 1/Bau 24, Erlangen 91058, Germany}
\affiliation{Faculty of Physics, M.V.Lomonosov Moscow State University, Moscow 119991, Russia}

\author{F.Ya.Khalili}
\affiliation{Faculty of Physics, M.V.Lomonosov Moscow State University, Moscow 119991, Russia}


\begin{abstract}
  Polarization quasi-probability distribution (PQPD) is defined in the Stokes space, and it enables the calculation of mean values and higher-order moments for polarization observables using simple algebraic averaging. It can be reconstructed with the help of polarization quantum tomography and provides a full description of the polarization properties of quantum states of light.

  We show here that, due to its definition in terms of the discrete-valued Stokes operators, polarization quasi-probability distribution has singularities and takes negative values at integer values of the Stokes observables. However, in experiments with `bright' many-photon states, the photon-number resolution is typically smeared due to the technical limitations of contemporary photodetectors. This results in a PQPD that is positive and regular even for such strongly nonclassical states as single-photon seeded squeezed vacuum.

  This problem can be solved by `highlighting' the quantum state, that is, by adding a strong coherent beam into the orthogonal polarization mode. This procedure bridges polarization quantum tomography with the Wigner-function tomography, while preserving the main advantage of the first one, namely, immunity to the common phase fluctuations in the light path. Thus, it provides a convenient method for the verification of bright nonclassical states of light, such as squeezed Fock states.
\end{abstract}

\maketitle

\section{Introduction}

During the last decade, non-classical states of light became a necessary tool in many physical experiments, most notably, very high precision measurements \cite{Nature_2011}, quantum computations, and quantum cryptography (see e.g., review papers \cite{Gisin_RMP_74_145_2002, Kok_RMP_79_135_2007, Hammerer_RMP_82_1041_2010} and references therein). Non-classical light will be also used in the emerging class of experiments aimed at the preparation of mechanical objects in non-Gaussian quantum states \cite{Zhang_PRA_68_013808_2003, 10a1KhDaMiMuYaCh}.

In all these experiments, some method of characterization and verification of the generated quantum state is required. The standard method for this is the quantum tomography \cite{Vogel_PRA_40_2847_1989, Raymer2004}, which allows one to restore the Wigner function \cite{Schleich2001} of the quantum state using the data acquired by a set of homodyne measurements. However, in many cases the practical implementation of this method could be difficult, in particular because it requires an additional local oscillator light source with the phase locked with the explored light. This requirement is especially hard to fulfill in the case of pulsed broadband light which is very typical in experiments with non-classical light.

This problem can be avoided by using the polarization tomography, which allows one to restore the quasi-probability distribution for the three Stokes operators of the two polarization modes of light ---  so called polarization quasi-probability distribution (PQPD) \cite{Bushev_OS_91_526_2001, Karassiov_JOB_4_S366_2002, Karassiov_LF_12_948_2002, Karassiov2004}. Evidently, it is not sensitive to the common phase of both polarization and therefore immune to the common phase fluctuations. Due to this very reason, it does not allow one to restore the full quantum state of the light, but only its so-called {\em polarization sector}.  However, in most cases, the polarization sector information is sufficient~\cite{Marquardt_PRL_99_220401_2007, Kanseri_PRA_85_022126_2012, Mueller_NJP_14_085002_2012}.

A distinctive feature of the PQPD, which it shares with the Wigner function, as well as with the classical probability distributions, is that its gives correct one-dimensional marginal distributions (in this particular case for the Stokes variables). Therefore, similar to the Wigner function, the PQPD represents the natural choice for the probability distributions in the classical hidden variables models. Expanding this analogy, it is possible to expect that PQPDs of `truly non-classical', {\it e.g.} non-Gaussian quantum states should demonstrate some non-trivial features, like negativity. However, as we show below, the discrete valued nature of the Stokes observables makes the situation a bit more complicated.

For optomechanical experiments, especially interesting are bright (with large mean number of photons) states, because they more effectively interact with mechanical objects (note that the masses of even most tiny nanobeams and nanomembranes used in these experiments are huge in comparison with the optical quanta `masses' $\hbar\omega/c^2\lesssim10^{-35}\,{\rm kg}$). For example, it was shown more than 30 years ago that the squeezed vacuum state allows to improve the sensitivity of optical interferometric displacement sensors \cite{Caves1981}. Recently, this idea was implemented in the laser interferometric gravitation-wave detector GEO-600 \cite{Nature_2011}. In a similar way, bright quantum non-Gaussian  states, like the squeezed single-photon state $\hat{\mathcal{S}}(r)\ket{1}$, where $\hat{\mathcal{S}}(r)$ is the squeezing operator, see Eq.\,\eqref{S_of_r}, are more attractive for the non-Gaussian optomechanics than their non-squeezed counterparts, for example the `ordinary' single-photon state $\ket{1}$, considered {\it e.g.} in Refs.\,\cite{Zhang_PRA_68_013808_2003, 10a1KhDaMiMuYaCh}.

Note that depending on the degree of squeezing $r$, the mean energy of a squeezed single-photon state can be arbitrary large. But independently of its mean energy, this state always possesses such essentially non-classical features as the negative-valued Wigner function and orthogonality to other squeezed Fock states $\hat{\mathcal{S}}(r)\ket{n\ne1}$ with the same degree of squeezing $r$.

The primary goal of this paper is to explore the applicability of the polarization tomography to the verification of bright non-Gaussian quantum states, and the second goal is to analyze the non-classical behavior of PQPD.

In Sec.\,\ref{sec:review}, we reproduce the basic formalism of the polarization tomography that could be found in the literature. In Sec.\,\ref{sec:loss} we discuss the effects of photodetectors' non-idealities and of the optical losses. In Sec.\,\ref{sec:LP}, which is devoted to the second goal, we consider linearly polarized light pulses and show, using this simple particular case, that the PQPD can be negative even for the states of light typically considered as essentially classical (like the coherent quantum state). We also discuss a possible experimental setup aimed at the demonstration of this negativity. In Sec.\,\ref{sec:P2W} we return to our primary goal and consider light containing some quantum state in one polarization and a coherent quantum state $\ket{\alpha_0}$ in the other one. It easy to see that if $|\alpha_0|\to\infty$ then the polarization tomography of this state reduces to ordinary tomography with the coherent quantum state serving as the local oscillator. We formulate requirements for the minimal value of $|\alpha_0|$ and for the photodetectors' parameters that are necessary to obtain the negative-valued PQPD in this setup. The Appendix contains some cumbersome calculations, which are not necessary for understanding the main results of this paper.

\section{PQPD and the polarization characteristic function}\label{sec:review}

Following the literature (see, {\it e.g.}, \cite{Bushev_OS_91_526_2001, Karassiov_LF_12_948_2002, Karassiov2004}), we introduce the polarization characteristic function as follows:
\begin{equation}\label{chi}
  \chi(u_1,u_2,u_3) := \Tr\left[\hat{\rho}\hat{\chi}(u_1,u_2,u_3)\right] ,
\end{equation}
where $\hat{\rho}$ is the density operator of a two-mode (horizontal and vertical polarizations) quantum state of light,
\begin{gather}
  \hat{\chi}(u_1,u_2,u_3)
  = \exp\biggl(i\sum_{i=1}^3u_i\hat{S}_i\biggr)
  = \exp\left[
        i(\hat{{\rm a}}_H^\dagger\ \hat{{\rm a}}_V^\dagger)\smatrix{u_1}{w^*}{w}{-u_1}
          \svector{\hat{{\rm a}}_H}{\hat{{\rm a}}_V}
      \right] , \label{hat_chi} \\
  w = u_2 + iu_3 \,,
\end{gather}
$\hat{{\rm a}}_H$, $\hat{{\rm a}}_V$ are the annihilation opearators for these modes,
\begin{align}
  & \hat{S}_1 = \hat{n}_H - \hat{n}_V \,, &
  & \hat{S}_2 = \hat{{\rm a}}_V^\dagger\hat{{\rm a}}_H
      + \hat{{\rm a}}_H^\dagger\hat{{\rm a}}_V \,, &
  & \hat{S}_3 = i(\hat{{\rm a}}_V^\dagger\hat{{\rm a}}_H
      - \hat{{\rm a}}_H^\dagger\hat{{\rm a}}_V)
\end{align}
are the Stokes operators, and
\begin{align}
  \hat{n}_H &= \hat{{\rm a}}_H^\dagger\hat{{\rm a}}_H \,, &
  \hat{n}_V &= \hat{{\rm a}}^\dagger_V\hat{{\rm a}}_V
\end{align}
are the photon-number operators in the $H,V$ modes. The PQPD is given by the Fourier transform of $\chi(u_1,u_2,u_3)$:
\begin{equation}\label{chi2W}
  W(S_1,S_2,S_3) = \intinfty\chi(u_1,u_2,u_3)
    \exp\biggl(-i\sum_{i=1}^3 u_iS_i\biggr)\,\frac{du_1du_2du_3}{(2\pi)^3}\,.
\end{equation}

An important feature of the Stokes operators, crucial for our consideration below, is that their eigenvalues are integer numbers varying from -$\infty$ to $\infty$. Therefore, the marginal characteristic functions $\bigl\langle\exp\bigl(iu_i\hat{S_i}\bigr)\bigr\rangle$ $(i=1,2,3)$ for these operators are $2\pi$-periodic in their argument, and the corresponding marginal probability distributions for $S_{1,2,3}$ are equal to sums of $\delta$-functions at the integer values of their arguments (we prefer to use the continuous-valued Fourier transformation here, which gives delta-functions instead of delta-symbols, for the sake of consistency with the treatment below).

\begin{figure}
  \includegraphics[width=0.48\textwidth]{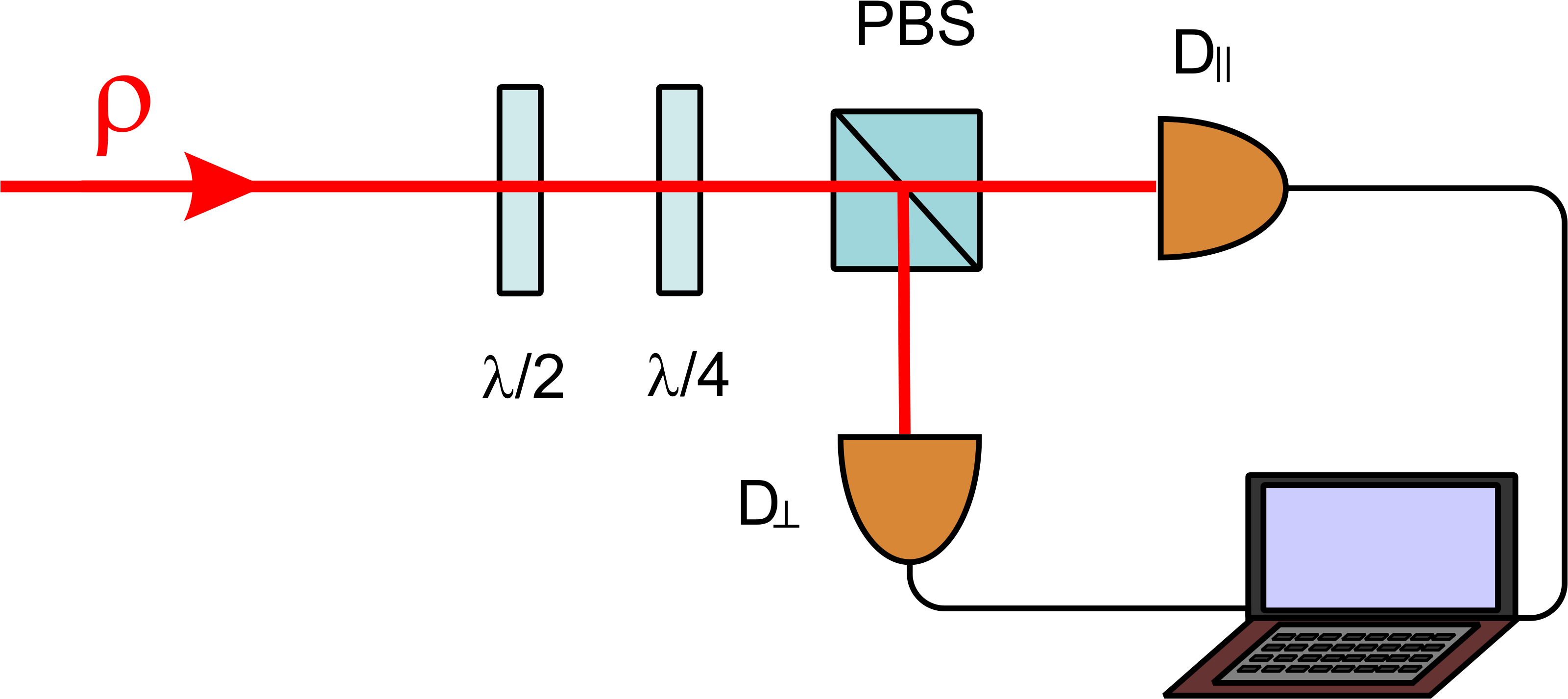}
  \caption{The setup for polarization tomography~\cite{Bushev_OS_91_526_2001,Karassiov2004}. PBS is the polarizing beam splitter, ${\rm D}_\|$ and ${\rm D}_\bot$ are the photodetectors. The signals from the detectors are processed by either digital or analog electronics, after which a computer calculates the probability distributions $W_{\theta\phi}(n)$ and performs the Radon transformation.}\label{fig:scheme}
\end{figure}

The characteristic function \eqref{chi} can be readily restored using the polarization tomography setup shown in Fig.\,\ref{fig:scheme}. This setup provides the probability distribution $W_{\theta\phi}(n)$ for the difference of the photon numbers in two orthogonal polarization modes measured by two photon counters ${\rm D}_\|$, ${\rm D}_\bot$:
\begin{multline}\label{S_theta_phi}
  \hat{S}_{\theta\phi}
  = \hat{{\rm a}}_\|^\dagger\hat{{\rm a}}_\|
    - \hat{{\rm a}}_\bot^\dagger\hat{{\rm a}}_\bot
  = (\hat{{\rm a}}_H^\dagger\ \hat{{\rm a}}_V^\dagger)
      \smatrix{\cos\theta}{e^{-i\phi}\sin\theta}{e^{i\phi}\sin\theta}{-\cos\theta}
      \svector{\hat{{\rm a}}_H}{\hat{{\rm a}}_V} \\
  = \hat{S}_1\cos\theta + (\hat{S}_2\cos\phi + \hat{S}_3\sin\phi)\sin\theta ,
\end{multline}
where
\begin{subequations}\label{aHV2a_det}
  \begin{align}
    \hat{{\rm a}}_\| &= \hat{{\rm a}}_H\cos\frac{\theta}{2}
      + \hat{{\rm a}}_Ve^{-i\phi}\sin\frac{\theta}{2}  \,,\\
    \hat{{\rm a}}_\bot &= \hat{{\rm a}}_H\sin\frac{\theta}{2}
      - \hat{{\rm a}}_V e^{-i\phi}\cos\frac{\theta}{2}
  \end{align}
\end{subequations}
are the annihilation operators for these modes and the angles $\theta$, $\phi$ depend on the orientations of the half- and quarter-wave plates shown in Fig.\,\ref{fig:scheme}. The characteristic function of this probability distribution is equal to
\begin{equation}\label{W2chi}
  \chi_{\theta\phi}(\lambda) = \sum_{n=-\infty}^{\infty}W_{\theta\phi}(n)e^{i\lambda n}
  = \Tr[\rho\hat{\chi}_{\theta\phi}(\lambda)] ,
\end{equation}
where
\begin{equation}\label{hat_chi_lambda}
  \hat{\chi}_{\theta\phi}(\lambda)
  = \exp\bigl[i\bigl(\lambda\hat{S}_{\theta\phi}\bigr)\bigr] .
\end{equation}
Comparing Eqs.\,\eqref{hat_chi} and \eqref{hat_chi_lambda}, it is easy to see that
\begin{equation}\label{chi2chi}
  \chi(u_1,u_2,u_3) = \chi_{\theta\phi}(\lambda) \,,
\end{equation}
with
\begin{align}\label{Radon}
  u_1 &= \lambda\cos\theta \,, & w = \lambda e^{i\phi}\sin\theta \,.
\end{align}

The chain of equalities (\ref{W2chi}, \ref{chi2chi}, \ref{chi2W}) forms, in essence, the Radon transformation which allows to calculate the PQPD from the experimentally acquired set of the distributions $W_{\theta\phi}(\lambda)$.

Taking into account that for any angle $\vartheta$,
\begin{equation}
  \hat{\mathcal{U}}^\dagger(\vartheta)\hat{S}_{1,2,3}\hat{\mathcal{U}}(\vartheta)
    \equiv \hat{S}_{1,2,3} \,,
\end{equation}
where
\begin{equation}
  \hat{\mathcal{U}}(\vartheta) = e^{-i\vartheta(\hat{n}_H + \hat{n}_V)}
\end{equation}
is the evolution operator which introduces a common phase shift $\vartheta$ into both polarizations, it is easy to see that the polarization characteristic function is invariant to this transformation:
\begin{equation}\label{chi_zeta}
  \Tr\left[
      \hat{\rho}\,
        \hat{\mathcal{U}}^\dagger(\vartheta)\hat{\chi}(u_1,u_2,u_3)\hat{\mathcal{U}}(\vartheta)
    \right]
  \equiv \Tr\left[\hat{\rho}\hat{\chi}(u_1,u_2,u_3)\right] .
\end{equation}
Therefore, the PQPD is not sensitive to any common (polarization-independent) fluctuations of the light optical path.

At the same time, it follows from Eq.\,\eqref{chi_zeta} that
\begin{equation}\label{chi_polar}
  \chi(u_1,u_2,u_3) = \Tr\left[\hat{\rho}_{\rm polar}\hat{\chi}(u_1,u_2,u_3)\right] ,
\end{equation}
where
\begin{equation}\label{rho_polar}
  \hat{\rho}_{\rm polar} = \int_{2\pi}
      \hat{\mathcal{U}}(\vartheta)\hat{\rho}\,\hat{\mathcal{U}}^\dagger(\vartheta)\,
    \frac{d\vartheta}{2\pi}\
  = \sum_{\substack{n_H,n_V = 0\\n_H',n_V' = 0}}^\infty
      \ket{n_Hn_V}\bra{n_Hn_V}\hat{\rho}\ket{n_H'n_V'}\bra{n_H'n_V'}
      \delta_{n_H+n_V\,n_H'+n_V'}
\end{equation}
is the polarization sector of the density operator equal to the incoherent sum of the `slices' of the density operator with given total numbers of quanta. Therefore, the polarization tomography restores only part of the light quantum state, namely, its polarization sector~\cite{Marquardt_PRL_99_220401_2007}.

\section{Quantum efficiency, optical losses, and photon-number integration}\label{sec:loss}

In the above consideration, it was assumed implicitly that the photodetectors are ideal and are able to exactly count all incident quanta. Their non-ideal quantum efficiency $\eta<1$ can be modeled by imaginary grey filters with the power transmissivity $\eta$, which mix the photodetectors input fields with some vacuum fields:
\begin{equation}\label{eff_loss}
  \hat{{\rm a}}_{\|,\bot}
  \to \sqrt{\eta}\,\hat{{\rm a}}_{\|,\bot} + \sqrt{1-\eta}\,\hat{{\rm b}}_{\|,\bot} \,,
\end{equation}
where $\hat{{\rm b}}_{\|,\bot}$  are the annihilation operators of the vacuum fields.

It is easy to show that these grey filters can be replaced by a single filter located at the input of the scheme of Fig.\,\ref{fig:scheme}, with some evident redefinition of the vacuum fields. This means that we can consider the photodetectors as ideal ones but take into account their non-unity quantum efficiency by introducing the corresponding effective losses into the incident light. Note that other optical losses can be also taken into account here by replacing the photodetectors quantum efficiency in Eq.\,\eqref{eff_loss} by the {\it unified quantum efficiency} of the scheme, equal to the probability for an incident photon to reach one of the photodetectors and be detected.

Another important shortcoming of contemporary photon-counting detectors is that their counting rate does not exceed $\sim10^7\,{\rm s}^{-1}$, which means that in the case of nanosecond and shorter pulses typically used in non-linear optics, they can count only one photon per pulse. More advanced transition-edge sensors can resolve up to $10$ photons, having at the same time high quantum efficiency, up to 95\%, but they are slow, difficult to use, and expensive \cite{Wildfeuer_PRA_80_043822_2009}.

In experiments with bright multi-photon pulses, {\it photon-number integrating} detectors are used instead, whose output signal is linearly proportional to the input number of quanta, but contaminated by additive noise. In the case of picosecond pulses used, {\it e.g.,} in  \cite{Hansen_OL_26_1714_2001, Kanseri_PRA_85_022126_2012}, this noise is equivalent to a measurement error of $\sigma\sim10^2$ quanta \cite{Stobinska_PRA_86_063823_2012}. Here we will model this noise by means of the Gaussian smoothing of the probability distribution $W_{\theta\phi}$:
\begin{equation}
  \tilde{W}_{\theta\phi}(y) = \sum_{n=0}^\infty \frac{W_{\theta\phi}(n)}
    {\sqrt{2\pi\sigma^2}}\,\exp\left[-\dfrac{(y-n)^2}{2\sigma^2}\right] \,.
\end{equation}
The corresponding smoothed characteristic function
\begin{equation}\label{chi_smooth}
  \tilde{\chi}(u_1,u_2,u_3) = \intinfty\tilde{W}_{\theta\phi}(y)e^{i\lambda y}\,dy
  = \chi(u_1,u_2,u_3)e^{-\sigma^2\lambda^2/2} \,,
\end{equation}
being substituted into Eq.\,\eqref{chi2W}, gives the smoothed PQPD:
\begin{equation}\label{tilde_chi2W}
  \tilde{W}(S_1,S_2,S_3) = \intinfty\tilde{\chi}(u_1,u_2,u_3)
    \exp\biggl(-i\sum_{i=1}^3 u_iS_i\biggr)\,\frac{du_1du_2du_3}{(2\pi)^3}\,.
\end{equation}

\section{Linearly polarized quantum states}\label{sec:LP}

To explore the negativity features of the PQPD, consider a simple particular case of linearly polarized quantum states, with only the $H$ mode excited and the $V$ mode in the vacuum state:
\begin{equation}\label{rho_LP}
  \hat{\rho} = \hat{\rho}_H\otimes\ket{0}_V\,{}_V\bra{0} \,.
\end{equation}
It follows from Eqs.\,(\ref{chi_polar}, \ref{rho_polar}) that in this case,
\begin{equation}\label{chi_LP}
  \chi(u_1,u_2,u_3) = \sum_{n=0}^\infty\rho_{H\,nn}\chi(u_1,u_2,u_3|n) \,,
\end{equation}
where
\begin{equation}
  \rho_{H\,nn} = \bra{n}\hat{\rho}_H\ket{n}
\end{equation}
and $\chi(u_1,u_2,u_3|n)$ is the characteristic function for the case of the $n$-photon Fock state in the $H$ mode; it was shown in paper \cite{Karassiov_LF_12_948_2002} that it is equal to
\begin{equation}
  \chi(u_1,u_2,u_3|n) = (\cos\lambda + iu_1\sinc\lambda)^n.
\end{equation}

The corresponding smoothed characteristic function, produced by photon-number integrating detectors, is equal to (assuming that $\sigma\gg1$ and, therefore, $\lambda\ll1$)
\begin{multline}
  \tilde{\chi}(u_1,u_2,u_3)
  \approx\sum_{n=0}^\infty\rho_{H\,nn}\left(1-\frac{\lambda^2}{2} + iu_1\right)^n
    e^{-\lambda^2\sigma^2/2} \\
  \approx\sum_{n=0}^\infty\rho_{H\,nn}
    \exp\left[-\frac{\sigma^2u_1^2}{2}  + inu_1 - \frac{(n+\sigma^2)|w|^2}{2}\right],
\end{multline}
and the smoothed PQPD [see Eq.\,\eqref{tilde_chi2W}] is equal to
\begin{equation}
  \tilde{W}(S_1,S_2,S_3)
  \approx\sum_{n=0}^\infty\frac{\rho_{H\,nn}}{(2\pi)^{3/2}\sigma(n+\sigma^2)}
      \exp\left[-\frac{(S_1-n)^2}{2\sigma^2} - \frac{S_{23}^2}{2(n+\sigma^2)}\right] ,
\end{equation}
where
\begin{equation}
  S_{23} = \sqrt{S_2^2 + S_3^2} \,.
\end{equation}
This result is completely intuitive and does not contain any `non-classical' features, like the negativity.

Consider, however, the exact not-smoothed PQPD. Unfortunately, the general equation for $W(S_1,S_2,S_3)$ in this case can not be expressed in any simple analytical form, but for our purposes, its marginal distributions are sufficient.

The marginal characteristic function for $S_1$ is given by
\begin{equation}
  \chi(u_1,0,0) = \sum_{n=0}^\infty\rho_{H\,nn}e^{iu_1n} \,,
\end{equation}
The corresponding marginal probability distribution,
\begin{equation}\label{W_1_LP}
  W_1(S_1) = \sum_{n=0}^\infty\rho_{H\,nn}\delta(S_1-n),
\end{equation}
is equal to the photon-number distribution for the state $\hat{\rho}_H$. The explanation is evident: the Stokes variable $S_1$ is equal to the difference of photon numbers in two polarizations, and in the case we consider here, the $V$ mode does not contain any quanta at all.

Much more interesting is the behavior of the Stokes variables $S_2,\,S_3$. Note that the characteristic function \eqref{chi_LP} does not depend on the angle $\phi$ and therefore the corresponding PQPD is invariant with respect to rotation in the ${S_2,\,S_3}$ plane. From classical point of view, this symmetry is incompatible with the above-mentioned discreteness of the marginal distributions for $S_2$ and $S_3$: this combination of features can not be manifested by any (positive-valued) probability distribution. However, it is completely feasible in the case of quantum quasi-probability distributions, which can have negative valued areas.

To analyze this feature in more detail, consider the two-dimensional marginal distribution for $S_2,\,S_3$, which in this particular case is equal to (see Appendix \ref{app:LPM}):
\begin{multline}\label{W_23_LP}
  W_{23}(S_2,S_3) = \intinfty W(S_1,S_2,S_3)\,dS_1
  = \intinfty\chi(0,u_2,u_3)e^{-iu_2S_2-iu_3S_3}\,\frac{du_2du_3}{(2\pi)^2} \\
  = \sum_{n=0}^\infty\frac{\rho_{H\,nn}}{2^n}
      \sum_{k=0}^n\frac{n!}{k!(n-k)!}\,w_{|2k-n|}(S_{23})\,,
\end{multline}
where
\begin{subequations}\label{w_m}
  \begin{gather}
    w_0(S_{23}) = \delta(S_2)\delta(S_3) \,, \\
    w_{m>0}(S_{23}) = \frac{1}{2\pi}\partd{}{S_{23}}\begin{cases}
        -\dfrac{S_{23}}{|m|\sqrt{m^2-S_{23}^2}}\,, & S_{23} < m \,, \\[2ex]
        0\,, & S_{23}\ge m \,.
      \end{cases}
  \end{gather}
\end{subequations}

\begin{figure*}
  \includegraphics[width=\textwidth]{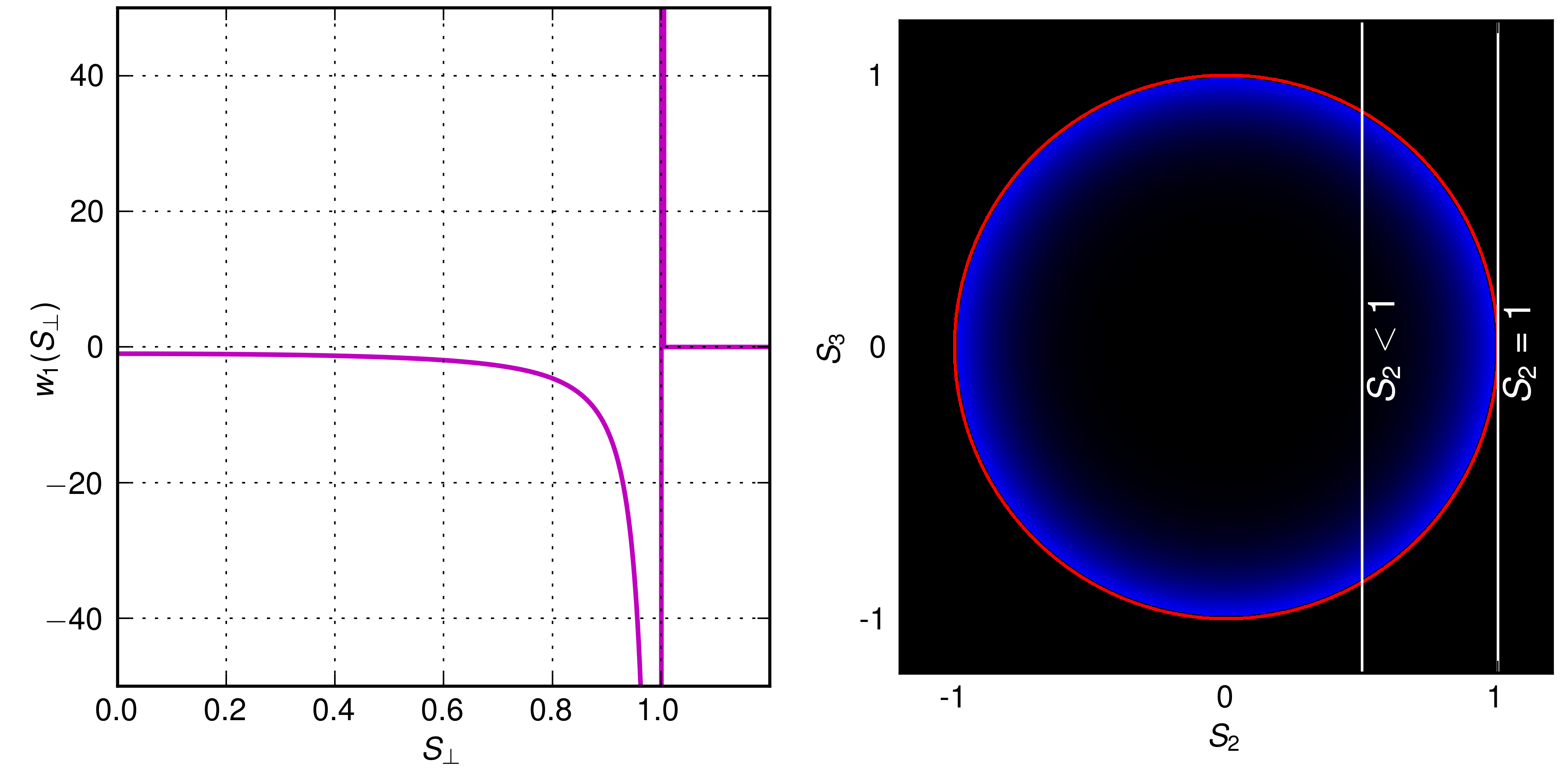}
  \caption{Left panel: plot of $w_1(S_{23})$. Right panel: color plot of $w_1(S_2,S_3)$ (blue: negative values, red: positive ones). Integration along the line $S_2=1$ gives infinity; integration along the lines $S_2<1$ gives zero due to the negative-valued areas.}\label{fig:W_LP}
\end{figure*}

The last equations, while looking a bit cumbersome, are actually very transparent. $W_{23}$ is equal to the weighted sum of functions $w_m$. The non-negative weight factors are given by the initial photon-number distribution convolved with the binomial distribution created by the beamsplitter. Each of the functions $w_m$, except for $w_0$, has negative values in the circular area $S_{23}<m$, see Fig.\,\ref{fig:W_LP} (left), where $w_1$ is plotted as the typical example. This means that the marginal distribution \eqref{W_23_LP} and therefore the corresponding PQPD $W(S_1,S_2,S_3)$ indeed has negative-valued areas for any quantum state $\hat{\rho}_H$.

It is this negativity that reconciles the discreteness of the marginal distributions and the rotation symmetry in ${S_2,\,S_3}$ plane, nullifying the marginal distributions for non-integer values of $S_{2,3}$. How it is possible is demonstrated by the right panel of Fig.\,\ref{fig:W_LP}, where the two-dimensional color plot of the function $w_1(S_2,S_3)$ is shown, with the positive-valued area of this function marked by red color and the negative-valued one by blue color. It is easy to see that integration along the line $S_2=1$ involves only positive values of $w_1(S_2,S_3)$ and thus gives a positive net value (actually infinity); and integration along the line $S_2<1$ involves both positive and negative values and thus can (and actually does) give zero. Due to the rotational symmetry of the picture, this result holds also for the marginal distribution of $S_3$, as well as of any combination $S_\phi = S_2\cos\phi + S_3\sin\phi$ (a similar result has been reported recently by A.V.Masalov \cite{Masalov_Klyshko8}).

This amazing structure of PQPD can be easily demonstrated experimentally using linearly polarized single-photon or even weak coherent light pulses. In the former case, with an account for the optical losses [see Eq.\,\eqref{eff_loss}],
\begin{equation}\label{p_0p_1}
  \bra{n}\hat{\rho}_H\ket{n} = p_0\delta_{n0} + p_1\delta_{n1} \,.
\end{equation}
where
\begin{align}
  & p_0 = 1-\eta \,, & & p_1 = \eta \,.
\end{align}
In the latter one, assuming that $\alpha\ll1$ and taking into account that the losses only decrease the mean number of quanta of the coherent state: $\alpha\to\alpha\sqrt{\eta}$, and it still remains coherent, we get the same equation \eqref{p_0p_1}, but with
\begin{align}
  & p_0 = e^{-|\alpha|^2} \approx 1-|\alpha|^2 \,, & & p_1 \approx |\alpha|^2 \,.
\end{align}

\begin{figure}
  \includegraphics[width=0.48\textwidth]{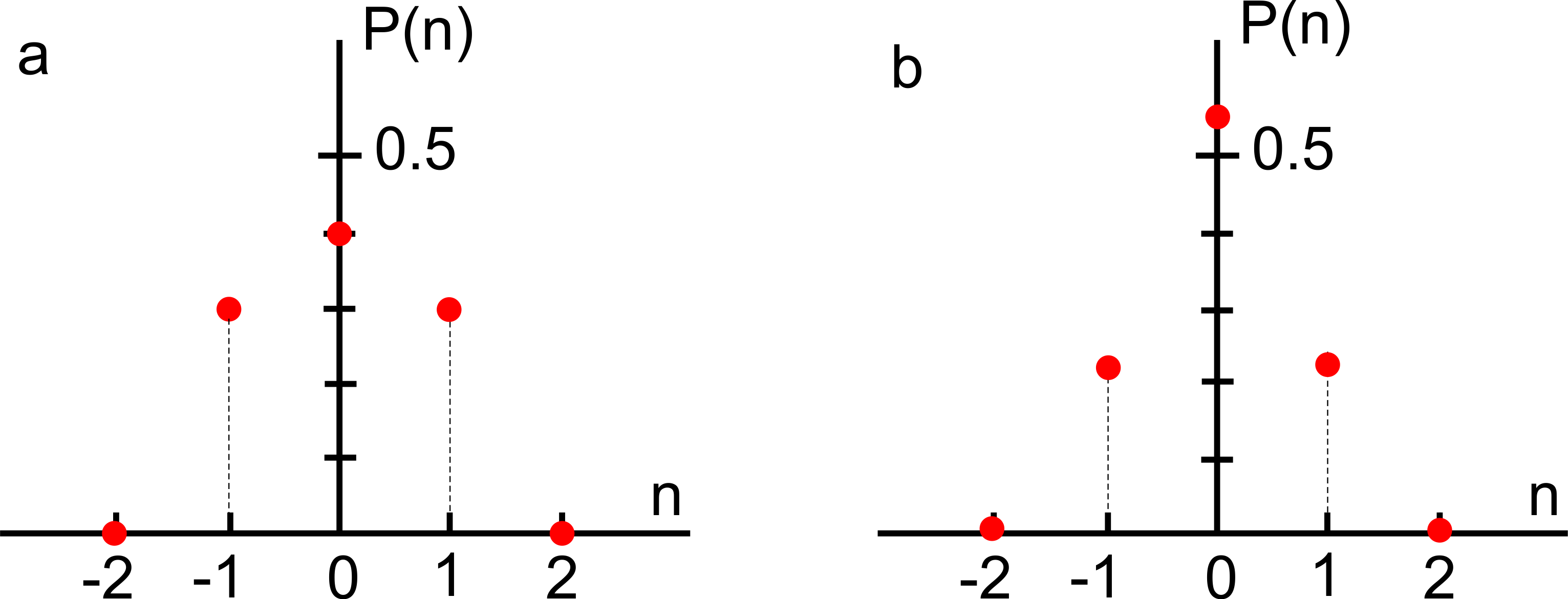}
  \caption{Typical probability distributions for a single-photon state (a) and a coherent state with $\alpha=1$ (b) at the input of the polarization tomography setup. The QE of the detectors is $\eta=0.6$ and the angle $\theta$ is chosen to be $\pi/2$.}\label{fig:dist_QE}
\end{figure}

In both these simple cases, in order to restore the marginal distribution \eqref{W_23_LP}, it is sufficient that the experimentalist measures only the distribution $W_{\theta\phi}$ for $\theta=\pi/2$ [see Eqs,\,\eqref{Radon}], which has a very simple form shown in Fig.\,\ref{fig:dist_QE}. Note that if the distributions are measured by a single-photon detector, no two-photon events will be observed. The presence of two-photon states in the density matrix $\rho_H$ (as in the case of a coherent state) will only increase the probability of a single-count event $p_1$ and reduce the probability of a no-count event $p_0$. The two-dimensional marginal Radon transformation \eqref{W_23_LP}, applied to this distribution, gives
\begin{equation}
  W_{23}(S_2, S_3) = p_0\delta(S_2)\delta(S_3) + p_1w_1(S_{23}),
\end{equation}
i.e., a $\delta$-function peak at $S_2=S_3=0$, surrounded by the negative-valued area provided by $w_1$.

At first sight it looks strange that PQPD can be negative valued even for such a `perfectly classical' state as the coherent one. However, it was emphasized {\it e.g.} in the review paper \cite{Braunstein2005} that classical local hidden-variable models require two necessary conditions: (i) the `classicality' of the quantum state, in the sense of positivity of its Wigner function, and (ii) the `classicality' of the measurement (only linear observables such as positions, momentums and their linear combinations have to be measured). The non-smoothed polarization tomography, which measures the discrete-valued Stokes variables, evidently violates the second assumption.

Another conclusion that can be derived from the above consideration is that the polarization tomography of linearly polarized light \eqref{rho_LP} can not be used to segregate the `classical' (with the Wigner function positive everywhere) quantum states $\hat{\rho}_H$ from `non-classical' ones, because in the smoothed case (with photon-number integrating detectors) it always gives positive PQPD, and in the non-smoothed case (with photon-number resolving detectors) it always gives PQPD with negativities (except of the trivial case of the vacuum state).

\section{`Highlighted' polarization quantum tomography}\label{sec:P2W}

The evident solution to this problem is the ``highlighting'' of the nonclassical features by feeding bright coherent light into the second polarization mode:
\begin{equation}\label{rho_alpha}
  \hat{\rho} = \hat{\rho}_H\otimes\ket{\alpha_0}_V\,{}_V\bra{\alpha_0}
\end{equation}
It is easy to see that in this case, the polarization tomography setup with fixed  $\theta=\pi/2$ exactly reproduces the ordinary quantum tomography setup, with the vertical polarization light serving as the local oscillator and the angle $\phi$ as the homodyne angle.

Indeed, consider the asymptotic case of a very strong coherent field, $|\alpha_0|\to\infty$. In this case, the operator $\hat{{\rm a}}_V$ in Eq.\,\eqref{hat_chi} can be replaced by its mean value $\alpha_0$, which gives the following equation for polarization characteristic function:
\begin{equation}\label{C_s2C_pol_asy}
  \chi(0,u_2,u_3) \approx \chi_s(\alpha_0w^*) \,,
\end{equation}
where
\begin{equation}\label{chi_s}
  \chi_s(z) = \Tr\bigl\{
    \hat{\rho}_H
      \exp\bigl[i\bigl(z\hat{{\rm a}}_H^\dagger + z^*\hat{{\rm a}}_H\bigr)\bigr]
    \bigr\}
\end{equation}
is the symmetrically ordered characteristic function for the state $\hat{\rho}_H$, whose Fourier transformation  gives the Wigner function for this state:
\begin{equation}\label{chi_s2W}
  W(x,p)
  = \intinfty\chi_s(z)\exp\bigl[-i\sqrt{2}(x\Re z + p\Im z)\bigr]\,\frac{d^2z}{2\pi^2} \,.
\end{equation}

A rigorous treatment of this problem (see Appendix \ref{app:RA}) shows that indeed a relation between the smoothed polarization characteristic function and the symmetrically ordered characteristic function exists, which in the reasonable particular case of not very bright quantum state $\rho_H$,
\begin{equation}\label{small_n}
  \mean{n} \ll \sigma^2 \,,
\end{equation}
where $\mean{n}$ is the mean number of quanta, simplifies to the smoothed version of Eq.\,\eqref{C_s2C_pol_asy}:
\begin{equation}\label{chi_s2chi}
  \tilde{\chi}(0,u_2,u_3) = \chi_s(\alpha_0w^*)e^{-\sigma^2|w|^2/2} \,.
\end{equation}

With an account for the optical losses [see the discussion around Eq.\,\eqref{eff_loss} and Appendix \ref{app:loss}], this equation takes the following form:
\begin{equation}\label{chi_s2chi_loss}
  \tilde{\chi}(0,u_2,u_3) = \chi_s(\zeta)e^{-\epsilon^2\zeta^2/2}
\end{equation}
where
\begin{equation}\label{zeta}
  \zeta = \zeta' + i\zeta'' = \sqrt{\eta}\,\alpha_0w^*
\end{equation}
and
\begin{equation}
  \epsilon^2 = \frac{1}{\eta}\left(1 - \eta + \frac{\sigma^2}{|\alpha_0|^2}\right) .
\end{equation}
is the total `quantum inefficiency' of the tomography scheme, which takes into account both the optical losses and the finite value of $\alpha_0$.

Finally, Fourier transformation of this equation gives the relation between the Wigner function and the smoothed PQPD:
\begin{equation}\label{W2C_pol_s}
  \tilde{W}_{23}(S_2,S_3) = \frac{1}{\pi\eta|\alpha_0|^2\epsilon^2}\intinfty W(x,p)
    \exp\biggl[
        -\frac{|S_2 - iS_3 - \sqrt{2\eta}\,\alpha_0^*(x+ip)|^2}
          {2\eta|\alpha_0|^2\epsilon^2}
      \biggr]
    dxdp \,.
\end{equation}
(compare with Eq.\,(7.35) of \cite{Raymer2004}). Note that in the ideal case of $\epsilon=0$, the Gaussian factor in this equation degenerates to the $\delta$-function, giving the exact one-by-one correspondence between $W_{23}(S_2,S_3)$ and $W(x,p)$.

Consider two examples of quantum states \eqref{rho_alpha}: a Gaussian squeezed vacuum state $\hat{\mathcal{S}}(r)\ket{0}_H$, and a non-Gaussian squeezed single-photon state $\hat{\mathcal{S}}(r)\ket{1}_H$, where
\begin{equation}\label{S_of_r}
  \hat{\mathcal{S}}(r)
  = \exp\left[\frac{r}{2}\left(\hat{{\rm a}}_H^\dagger{}^2-\hat{{\rm a}}_H^2\right)\right]
\end{equation}
is the squeezing operator.

In the first case,
\begin{equation}\label{chi_s_sqz0}
  \chi_s(z) = \exp\left(-\frac{z'{}^2e^{2r} + z''{}^2e^{-2r}}{2}\right) .
\end{equation}
It is shown in App.\,\ref{app:sqz0} that the corresponding smoothed marginal polarization characteristic function is equal to
\begin{equation}\label{C_pol_sqz0s}
  \tilde{\chi}(0,u_2,u_3)
  = \exp\left(-\frac{\delta_+^2\zeta'{}^2 + \delta_-^2\zeta''{}^2}{2}\right) ,
\end{equation}
where
\begin{equation}
  \delta_\pm^2 = e^{\pm2r} + \epsilon^2 \,.
\end{equation}
[it is easy to see that is can be obtained simply by substitution of Eq.\,\eqref{chi_s_sqz0} into \eqref{chi_s2chi}; however, the direct calculation of App.\,\ref{app:sqz0} allows one to formulate the explicit analog of condition \eqref{small_n} for this particular case].

Using then Eq.\,\eqref{chi2W}, we obtain the marginal PQPD that is Gaussian and thus  positive everywhere:
\begin{equation}\label{W23_S0d}
  \tilde{W}_{23}(S_2,S_3)
  = \frac{1}{2\pi\eta|\alpha_0|^2\delta_+\delta_-}\exp\left[
        -\frac{1}{2}\left(\frac{s_2^2}{\delta_+^2}+ \frac{s_3^2}{\delta_-^2}\right)
      \right] ,
\end{equation}
where
\begin{align}\label{S_norm}
  & s_2 = \Re\frac{S_2-iS_2}{\sqrt{\eta}\alpha_0^*} \,, &
  & s_3 = \Im\frac{S_2-iS_2}{\sqrt{\eta}\alpha_0^*}
\end{align}
are the normalized Stokes variables.

In the case of the squeezed single-photon state,
\begin{equation}\label{chi_s_sqz1}
  \chi_s(z) = \left(1 - z'{}^2e^{2r} - z''{}^2e^{-2r}\right)
    \exp\left(-\frac{z'{}^2e^{2r} + z''{}^2e^{-2r}}{2}\right) .
\end{equation}
It is shown in App.\,\ref{app:sqz1} that the corresponding smoothed marginal polarization characteristic function is equal to
\begin{equation}\label{C_pol_sqz1s}
  \tilde{\chi}(0,u_2,u_3) = \left(1 - \zeta'{}^2e^{2r} - \zeta''{}^2e^{-2r}\right)
    \exp\left(-\frac{\delta_+^2\zeta'{}^2 + \delta_-^2\zeta''{}^2}{2}\right) ,
\end{equation}
and correspondingly [using again Eq.\,\eqref{chi2W}],
\begin{equation}\label{tilde_W_23}
  \tilde{W}_{23}(S_2,S_3) = \frac{1}{2\pi\eta|\alpha_0|^2\delta_+\delta_-}
    \left(
        \frac{s_2^2e^{2r}}{\delta_+^4} + \frac{s_3^2e^{-2r}}{\delta_-^4}
        + \frac{\epsilon^4 -1}{\delta_+^2\delta_-^2}
      \right)
    \exp\left[
        -\frac{1}{2}\left(\frac{s_2^2}{\delta_+^2}+ \frac{s_3^2}{\delta_-^2}\right)
      \right] .
\end{equation}

It is easy to see that if
\begin{equation}\label{small_eps}
  \epsilon < 1  \,,
\end{equation}
that is if the photon-number integration, given by $\sigma$, is not very strong, and the quantum efficiency $\eta$ is sufficiently high, then the PQPD manifests negativity, caused of course by the negativity of the Wigner function. Note that in particular, the condition \eqref{small_eps} requires that the unified quantum efficiency of the scheme has to be higher than $1/2$ \cite{Raymer2004}.

However, for the negativity of the PQPD to be experimentally detectable it is important that the negative part is pronounced compared to the positive part. This imposes a requirement that $\epsilon$ should be smaller than a certain value, which strongly depends on the squeezing.

\begin{figure*}
  \includegraphics[width=\textwidth]{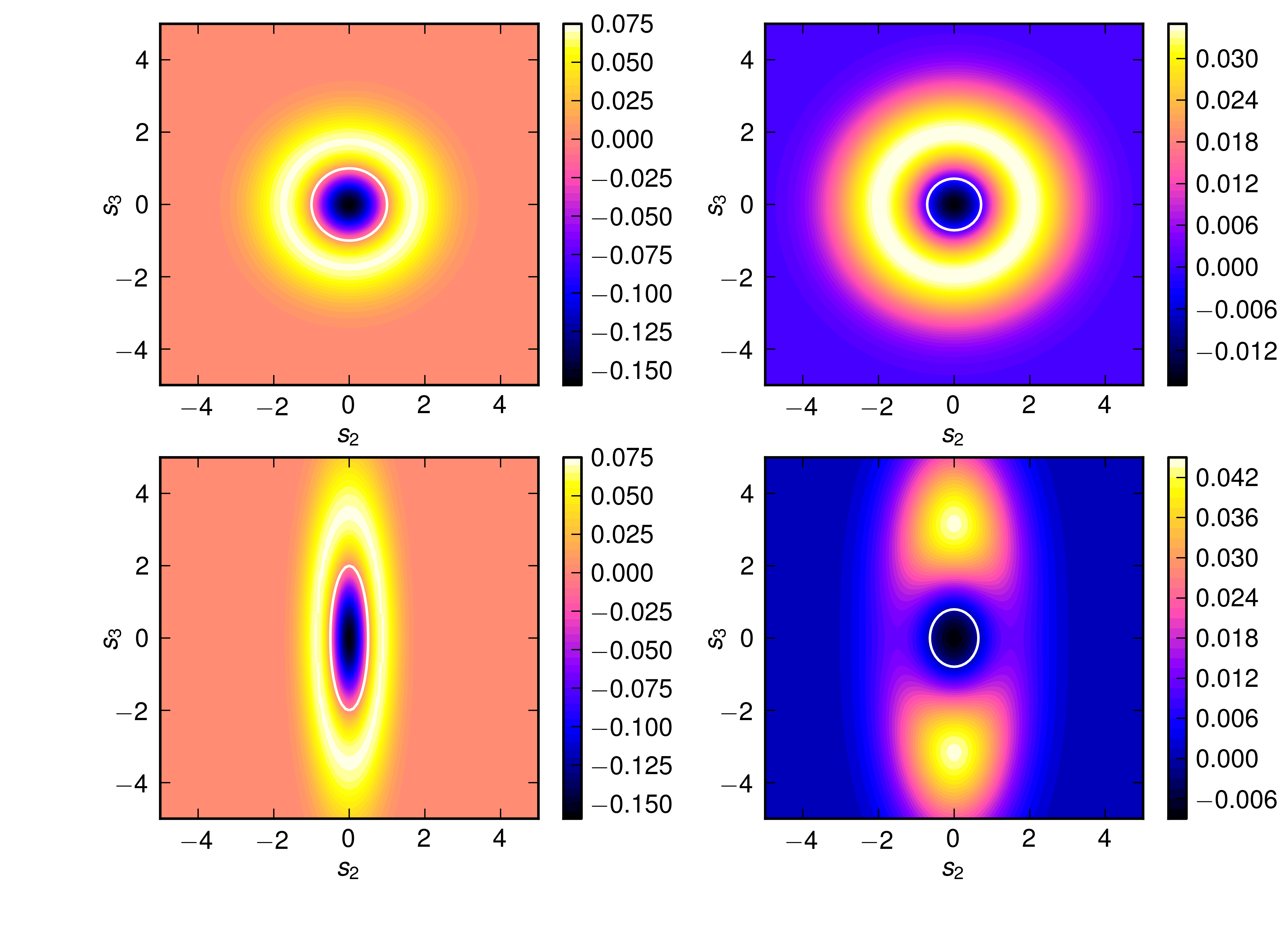}
  \caption{Contour plots of the quasi-probability distribution \eqref{tilde_W_23} as a function of the normalized Stokes parameters \eqref{S_norm} for the squeezed single-photon state in the absence of losses and photon-number integration (left column) and with $\epsilon^2=0.7$ (right column). Top row: no squeezing ($e^r=1$); Bottom row: 6\,db squeezing ($e^r=2$). The negative-valued areas are encircled by the white lines (the color corresponding to $W_{23}=0$ varies due to the different ratios of the maximal and the minimal values of $W_{23}$).}\label{fig:highlighted}
\end{figure*}

In Fig\,\ref{fig:highlighted}, the probability distribution \eqref{tilde_W_23} is plotted for the ordinary (non-squeezed) single-photon state and for the 6-db squeezed one. The left two plots correspond to the ideal case of $\epsilon=0$, the right ones, to the typical case of $\epsilon^2=0.7$. It can be seen from these plots that the negative-valued area of $\tilde{W}_{23}$ shrinks due to the losses but is only weakly affected by the squeezing. However, the depth of this area decreases very significantly in the squeezed case, due to the well known feature of vulnerability of the squeezing to the optical losses.

\begin{figure}
  \includegraphics[width=0.49\textwidth]{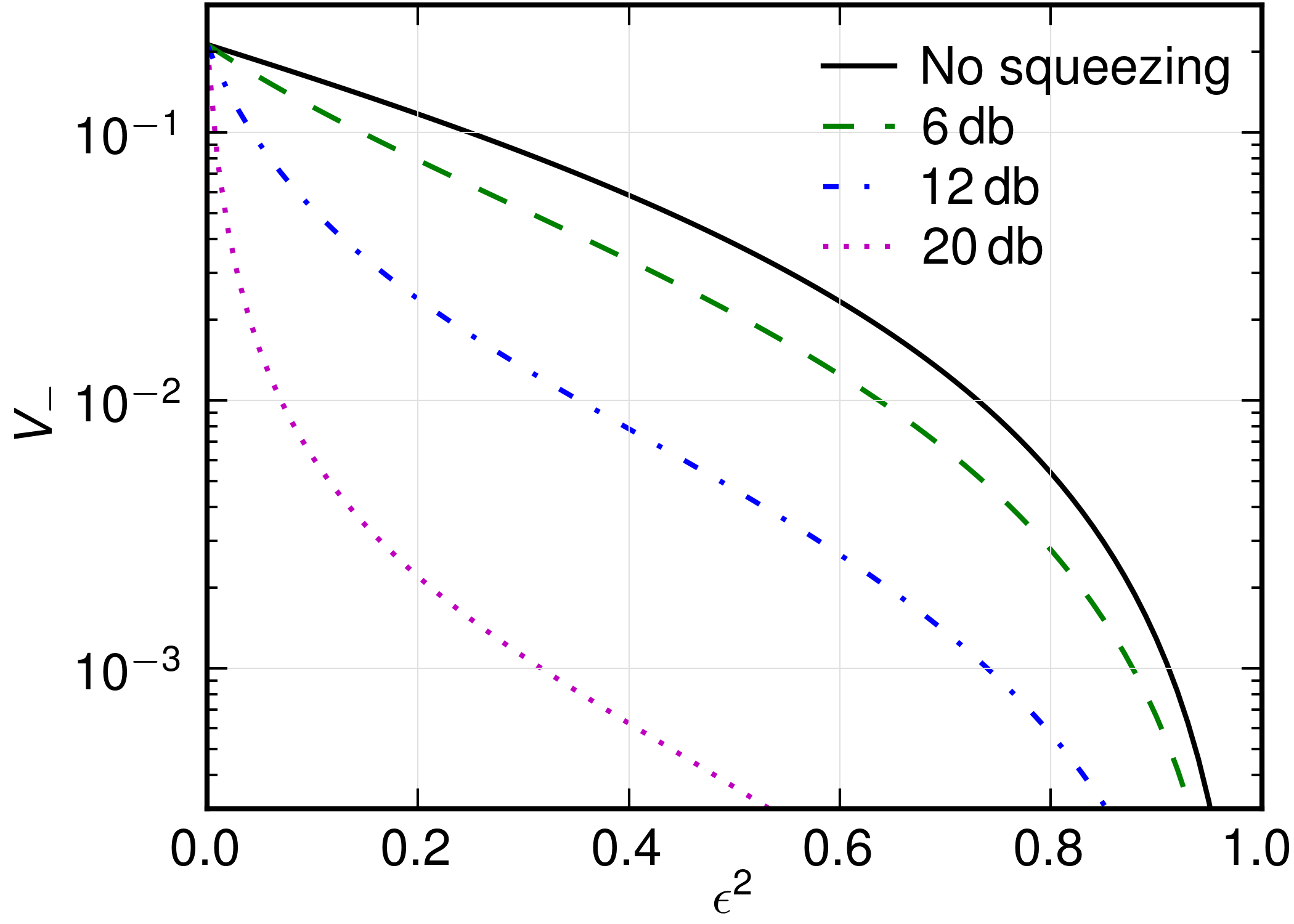}
  \caption{The volume of the negative-valued area of the quasi-probability distribution \eqref{tilde_W_23} as a function of the total quantum inefficiency $\epsilon^2$}\label{neg_vol}
\end{figure}

The convenient quantitative measure of the negativity, which takes both these effects into account, is the volume of the negative-valued part of the quasi-probability distribution:
\begin{equation}
  V_- = -\int_{\tilde{W}_{23}<0}\tilde{W}_{23}(S_2,S_3)dS_2dS_3 \,.
\end{equation}
It is plotted in Fig.\,\ref{neg_vol} as a function of $\epsilon^2$ for several values of the squeezing factor. It follows from this plot that unfortunately, for reasonable losses $\epsilon^2\gtrsim0.5$, only quite modest squeezing about 10\,db can be used. In order to use bright strongly squeezed states, the optical losses have to be reduced significantly, down to $\epsilon^2\lesssim0.1$

\section{Conclusion}

Thus, we have shown that the polarization quantum tomography is an essentially discrete-variable technique. It is aimed at finding the quasi-probability distribution of the Stokes observables whose quantum counterparts, the Stokes operators, have discrete spectra. In its rigorous version, polarization quantum tomography should involve measurements with photon-number resolving detectors leading to discrete experimental probability distributions. In this case, the reconstructed PQPD will contain nonclassical features, such as negativity areas, even for perfectly `classical' states. This demonstrates the connection between two standard signs of non-classicality: the discreteness of photon numbers and the negativity of quasi-probability distributions.

However, in an experiment with `bright' multi-photon states, it is usually impossible to perform measurements with single-photon resolution. Photon-number integration leads to the smearing of the probability distribution and therefore can prevent the observation of PQPD negativity even for some `very nonclassical' states such as the Fock ones.

This problem can be solved by `highlighting' the quantum state, that is, by adding a strong coherent beam into the orthogonal polarization mode. This procedure actually bridges polarization quantum tomography with the Wigner-function tomography; in the very strong highlighting case the former one simply reduces to the latter one. The negativity of the Wigner function will then be manifested in the negativity of the PQPD, provided that the losses are not too high and the photon-number integration is not too broad. This way one can test for nonclassicality bright quantum states of light, such as squeezed Fock states.

\acknowledgments

The work of M.Ch. was supported in part by the grant for the NATO project EAP.SFPP 984397 ``Secure Communication Using Quantum Information Systems''. The work of F.Kh. was supported by a grant from Dr. Hertha u. Helmut Schmauser-Stiftung, LIGO NSF grant PHY-0967049 and Russian Foundation for Basic Research grant No.11-02-00383-a.

\appendix

\section{Derivation of the marginal PQPD \eqref{W_23_LP}}\label{app:LPM}

Setting in \eqref{chi_LP} $u_1=u_2=0$, we get
\begin{equation}
  \chi(0,w) = \sum_{n=0}^\infty\rho_{H\,nn}\cos^n|w| \,.
\end{equation}
Therefore,
\begin{multline}
  W_{23}(S_2,S_3)
  = \frac{1}{(2\pi)^2}\sum_{n=0}^\infty\rho_{H\,nn}
      \int_{2\pi}d\varphi\intOinfty|w|d|w|\,e^{-iS_{23}|w|\cos\varphi}\cos^n\!|w| \\
  = \sum_{n=0}^\infty\frac{\rho_{H\,nn}}{2^n}\sum_{k=0}^n\frac{n!}{k!(n-k)!}\,
      W_{2k-n}(S_{23})
  = \sum_{n=0}^\infty\frac{\rho_{H\,nn}}{2^n}\sum_{k=0}^n\frac{n!}{k!(n-k)!}\,
      w_{|2k-n|}(S_{23})\,,
\end{multline}
where
\begin{gather}
  S_{23} = \sqrt{S_2^2 + S_3^2} \,, \qquad \varphi = \arg(S_2 + iS_2) \,, \\
  w_m(S_{23}) = \frac{W_m(S_{23}) + W_{-m}(S_{23})}{2} \,, \\
  W_m(S_{23}) = \frac{1}{(2\pi)^2}
      \int_{2\pi}d\varphi\intOinfty|w|d|w|\,e^{-i(S_{23}\cos\varphi + m)|w|}
  = \partd{F_m(S_{23})}{S_{23}} \,,
\end{gather}
{\allowdisplaybreaks\begin{multline}
  F_m(S_{23}) = \frac{i}{(2\pi)^2}\int_{2\pi}d\varphi\intOinfty d|w|\,
      \frac{e^{-i(S_{23}\cos\varphi + m)|w|}}{\cos\varphi} \\
  = \frac{i}{(2\pi)^2}\lim_{\gamma\to0}\int_{2\pi}d\varphi\intOinfty d|w|\,
      \frac{e^{-[\gamma + i(S_{23}\cos\varphi + m)]|w|}}{\cos\varphi} \\
  = \frac{1}{2\pi}\begin{cases}
      \displaystyle\lim_{\gamma\to0}\frac{\gamma}{(S_{23}^2+\gamma^2)^{3/2}}\,, & m=0\,,
        \\[2ex]
      -\dfrac{S_{23}}{|m|\sqrt{m^2-S_{23}^2}}\,, & m>0\text{ and }S_{23} < m \,, \\[2ex]
      0\,, & S_{23}\ge m > 0 \,,
    \end{cases}
\end{multline}}
which gives Eq.\,\eqref{w_m}.

\section{Polarization characteristic function of quantum states \eqref{rho_alpha}}\label{app:RA}

Consider the polarization characteristic function \eqref{chi} for the two-mode coherent state  $\ket{\alpha}_H\ket{\alpha_0}_V$, which was calculated in \cite{Karassiov_LF_12_948_2002}:
\begin{equation}
  \chi(u_1,u_2,u_3|\alpha)
  = \exp\bigl[
        -\varkappa|\alpha|^2 - \varkappa^*|\alpha_0|^2
        + i(\alpha\alpha_0^*w + \alpha^*\alpha_0w^*)\sinc\lambda
      \bigr] ,
\end{equation}
where
\begin{equation}
  \varkappa = 1 -\cos\lambda - iu_1\sinc\lambda \,.
\end{equation}
Expressing the density operator $\hat{\rho}_H$ through the Glauber's P-function,
\begin{equation}
  \hat{\rho}_H = \int\ket{\alpha}P(\alpha)\bra{\alpha}\,d^2\alpha \,,
\end{equation}
and using the well-known relations between $P(\alpha)$, the corresponding normally-ordered characteristic function $\chi_n(z)$, and the symmetric characteristic function \eqref{chi_s},
\begin{gather}
  \chi_n(z) = \Tr\bigl(\hat{\rho} e^{iz\hat{{\rm a}}^\dagger}e^{iz^*\hat{{\rm a}}}\bigr)
  = \int P(\alpha)e^{iz^*\alpha +z\alpha^*}\,d^2\alpha \,, \\
  \chi_s(z) = \chi_n(z)e^{-|z|^2/2} \,,
\end{gather}
we get the polarization characteristic function for an arbitrary quantum state of the form \eqref{rho_alpha}:
{\allowdisplaybreaks\begin{multline}
  \chi(u_1,u_2,u_3) = \int P(\alpha)\chi(u_1,u_2,u_3|\alpha)\,d^2\alpha \\
  = \frac{1}{\pi^2}\intinfty\chi_n(z)\exp\bigl[
        -\varkappa|\alpha|^2 + i(\alpha\alpha_0^*w + \alpha^*\alpha_0w^*)\sinc\lambda
          - \varkappa^*|\alpha_0|^2 - i(z^*\alpha + z\alpha^*)
      \bigr]d^2\alpha d^2z \\
  = \frac{1}{\pi\varkappa}\intinfty\chi_s(z)\exp\left(
        \frac{|z|^2}{2} - \frac{|z - \alpha_0w^*\sinc\lambda|^2}{\varkappa}
        - \varkappa^*|\alpha_0|^2
      \right)d^2z \,.
\end{multline}}
For our consideration below, we only need the part of this characteristic function with $u_1=0$:
\begin{equation}\label{C_s2C_pol_w}
  \chi(0,u_2,u_3) = \frac{1}{2\pi\sin^2\dfrac{|w|}{2}}\intinfty \chi_s(z)\exp\biggl(
      -\frac{1}{2}\cot^2\frac{|w|}{2}
      \left|z-\frac{2\alpha_0w^*}{|w|}\tan\frac{|w|}{2}\right|^2
    \biggr)
  d^2z \,.
\end{equation}
Smoothing this characteristic function [see Eq.\,\eqref{chi_smooth}] and taking into account that if $\sigma\gg1$ then only small values of $|w|\ll1$ are of relevance, we get:
\begin{equation}
  \tilde{\chi}(0,u_2,u_3)
  = \frac{2e^{-\sigma^2|w|^2/2}}{\pi|w|^2}
      \intinfty\chi_s(z)\exp\biggl(-\frac{2|z - \alpha_0 w^*|^2}{|w|^2}\biggr)d^2z \,.
\end{equation}
In the particular case of \eqref{small_n}, which is equivalent to the condition $|z|\gg|w|$, the Gaussian function in this equation can be approximated by the delta-function:
\begin{equation}
  \frac{2}{\pi|w|^2}\exp\biggl(-\frac{2|z - \alpha_0 w^*|^2}{|w|^2}\biggr)
    \to\delta(z - \alpha_0w^*),
\end{equation}
which gives Eq.\,\eqref{chi_s2chi}.

\section{Optical losses}\label{app:loss}

Let us start with the symmetrically ordered characteristic function of some quantum state $\hat{\rho}$:
\begin{equation}
  \chi_s(z) = \Tr\bigl\{
      \hat{\rho}\exp\bigl[i\bigl(z\hat{{\rm a}}^\dagger + z^*\hat{{\rm a}}\bigr)\bigr]
    \bigr\} .
\end{equation}
Using the description of the optical losses by means of an imaginary grey filter [see Eq.\,\eqref{eff_loss}], the characteristic function of the lossy optical mode can be expressed as follows:
\begin{multline}\label{chi_loss}
  \chi_s^{\rm loss}(z) = \Tr\bigl\{
      \hat{\mathcal{A}}\hat{\rho}\otimes\ket{0}_L\,{}_L\bra{0}\hat{\mathcal{A}}^\dagger
        \exp\bigl[i\bigl(z\hat{{\rm a}}^\dagger + z^*\hat{{\rm a}}\bigr)\bigr]
    \bigr\} \\
  = \Tr\bigl\{
      \hat{\rho}\exp\bigl[
          i\sqrt{\eta}\bigl(z\hat{{\rm a}}^\dagger + z^*\hat{{\rm a}}\bigr)
        \bigr]
      \bigr\}
      \times{}_L\bra{0}\exp\bigl[
            i\sqrt{1-\eta}\bigl(z\hat{{\rm b}}^\dagger + z^*\hat{{\rm b}}\bigr)
          \bigr]
        \ket{0}_L \\
  = \chi_s(\sqrt{\eta}\,z)e^{-(1-\eta)|z|^2/2} \,.
\end{multline}
where $\hat{\rho}$ is the initial density operator (before passing the light through the grey filter), $\ket{0}_L$ is the ground state of the ``losses'' (vacuum) mode, and $\hat{\mathcal{A}}$ is the unitary evolution operator corresponding to the transformation \eqref{eff_loss}. Substitution of this characteristic function into Eq.\,\eqref{chi_s2chi} gives Eq.\,\eqref{chi_s2chi_loss}.

\section{Smoothed PQPDs for the damped squeezed vacuum and squeezed single-photon states}
\label{app:sqz}

\subsection{Squeezed vacuum state}\label{app:sqz0}

Symmetrically ordered characteristic function for the squeezed vacuum state has the form
\begin{equation}
  \chi_s^{\rm loss}(z)
  = \exp\left(-\frac{\Delta_+^2z'{}^2 + \Delta_-^2z''{}^2}{2}\right) ,
\end{equation}
where
\begin{equation}
  \Delta_\pm^2 = \eta e^{\pm2r} + 1-\eta \,.
\end{equation}
Substitution of this characteristic function into Eq.\,\eqref{C_s2C_pol_w} gives:
\begin{equation}\label{C_pol_sqz0}
  \chi(0,u_2,u_3) = C_0\exp\left\{
      -\frac{2}{|w|^2}\left[
          \dfrac{\Delta_+^2}{\varkappa_+^2}\,\Re^2(\alpha_0w^*)
          + \dfrac{\Delta_-^2}{\varkappa_-^2}\,\Re^2(\alpha_0w^*)
        \right]
    \right\} ,
\end{equation}
where
\begin{gather}
  \varkappa_\pm^2 = \Delta_\pm^2 + \cot^2\frac{|w|}{2} \,, \\
  C_0 = \frac{1}{\varkappa_+\varkappa_-\sin^2\dfrac{|w|}{2}} \,.
\end{gather}

Suppose that the squeezing is not very strong (compare with Eq.\,\eqref{small_n}):
\begin{equation}\label{small_n_sqz}
  \Delta_+ \ll \frac{1}{|w|} \sim \sigma \,.
\end{equation}
In this case, smoothing of \eqref{C_pol_sqz0} gives Eq.\,\eqref{C_pol_sqz0s}.

\subsection{Squeezed single-photon state}\label{app:sqz1}

Using Eqs.\,(\ref{chi_s_sqz1}, \ref{chi_loss}), we get:
\begin{equation}
  \chi_s^{\rm loss}(z) = \left[1 - \eta\left(z'{}^2e^{2r} - z''{}^2e^{-2r}\right)\right]
    \exp\left(-\frac{\Delta_+^2z'{}^2 + \Delta_-^2z''{}^2}{2}\right) .
\end{equation}
Substitution of this characteristic function into Eq.\,\eqref{C_s2C_pol_w} gives:
\begin{multline}
  \chi(0,u_2,u_3) = C_0\left\{
      C_0^2(\eta\cos|w| + 1 - \eta)
      - \frac{4\eta}{|w|^2}\cot^2\frac{|w|}{2}\left[
          \dfrac{\Re^2(\alpha_0w^*)}{\varkappa_+^4}\,e^{2r}
          + \dfrac{\Im^2(\alpha_0w^*)}{\varkappa_-^4}\,e^{-2r}
        \right]
    \right\} \\ \times
    \exp\left\{
        -\frac{2}{|w|^2}\left[
            \dfrac{\Delta_+^2}{\varkappa_+^2}\,\Re^2(\alpha_0w^*)
            + \dfrac{\Delta_-^2}{\varkappa_-^2}\,\Re^2(\alpha_0w^*)
          \right]
      \right\}
\end{multline}
In the smoothed case of \eqref{small_n_sqz}, this equation simplifies to Eq.\,\eqref{C_pol_sqz1s}.


\begin{thebibliography}{10}

\bibitem{Nature_2011}
{J.Abadie {\it et al}},
\newblock Nature Physics {\bf 7}, 962 (2011).

\bibitem{Gisin_RMP_74_145_2002}
{N.~Gisin, G.~Ribordy, W.~Tittel, and H.~Zbinden},
\newblock Rev. Mod. Phys. {\bf 74}, 145 (2002).

\bibitem{Kok_RMP_79_135_2007}
{P.~Kok, W.~J.~Munro, K.~Nemoto, T.~C.~Ralph, J.~P.~Dowling, G.~J.~Milburn,}
\newblock Rev. Mod. Phys. {\bf 79}, 135 (2007).

\bibitem{Hammerer_RMP_82_1041_2010}
{K.~Hammerer, A.~S. S\o{}rensen, and E.~S. Polzik},
\newblock Rev. Mod. Phys. {\bf 82}, 1041 (2010).

\bibitem{Zhang_PRA_68_013808_2003}
{J.~Zhang, K.~Peng, and S.~L. Braunstein},
\newblock Phys. Rev. A {\bf 68}, 013808 (2003).

\bibitem{10a1KhDaMiMuYaCh}
{F.~Khalili, S.~Danilishin, H.~Miao, H.~Muller-Ebhardt, H.~Yang, and and Y.~Chen,}
\newblock Phys. Rev. Lett. {\bf 105}, 070403 (2010).

\bibitem{Vogel_PRA_40_2847_1989}
{K.~Vogel and H.~Risken},
\newblock Phys. Rev. A {\bf 40}, 2847 (1989).

\bibitem{Raymer2004}
M.G.Raymer, M.Beck, Lect. Notes Phys. {\bf 649}, 235 (2004).

\bibitem{Schleich2001}
{W. Schleich},
\newblock {\em Quantum Optics in Phase Space},
\newblock WILEY-VCH, Berlin, 2001.

\bibitem{Bushev_OS_91_526_2001}
{P.A Bushev, V.P.Karassiov, A.V.Masalov, and A.A.Putilin},
\newblock Optics and Spectroscopy {\bf 91}, 526 (2001).

\bibitem{Karassiov_JOB_4_S366_2002}
{V.P.Karassiov and A.V.Masalov},
\newblock J. Opt. B: Quantum Semiclass. Opt. {\bf 4}, S366 (2002).

\bibitem{Karassiov_LF_12_948_2002}
{V.P.Karassiov and A.V.Masalov},
\newblock Laser Physics {\bf 12}, 948 (2002).

\bibitem{Karassiov2004}
{V.P.Karassiov and A.V.Masalov},
\newblock {Journal of Experimental and Theoretical Physics} {\bf 99}, 51
  (2004).

\bibitem{Marquardt_PRL_99_220401_2007}
{Ch.~Marquardt, J.~Heersink, R.~Dong, M.~V.~Chekhova, A.~B.~Klimov, L.~L.~Sanchez-Soto,
U.~L.~Andersen, and G.~Leuchs,}
\newblock Phys. Rev. Lett. {\bf 99}, 220401 (2007).

\bibitem{Kanseri_PRA_85_022126_2012}
{B.Kanseri, T.Iskhakov, I.Agafonov, M.Chekhova, and G.Leuchs},
\newblock Phys. Rev. A {\bf 85}, 022126 (2012).

\bibitem{Mueller_NJP_14_085002_2012}
{C.~R.~M\"uller, B.~Stoklasa, C.~Peuntinger, C.~Gabriel, J.~Rehacek, Z.~Hradil,
A.~B.~Klimov, G.~Leuchs, Ch.~Marquardt, and L.~L.~Sanchez-Soto},
\newblock New Journal of Physics {\bf 14}, 085002 (2012).

\bibitem{Caves1981}
{C.M.Caves},
\newblock Physical Review D {\bf 23}, 1693 (1981).

\bibitem{Wildfeuer_PRA_80_043822_2009}
{C.~F.~Wildfeuer, A.~J.~Pearlman, J.~Chen, J.~Fan, A.~Migdall, and J.~P.~Dowling},
\newblock Phys. Rev. A {\bf 80}, 043822 (2009).

\bibitem{Hansen_OL_26_1714_2001}
{H.~Hansen, T.~Aichele, C.~Hettich, P.~Lodahl, A.~I.~Lvovsky, J.~Mlynek, and S.~Schiller,}
\newblock Opt. Lett. {\bf 26}, 1714 (2001).

\bibitem{Stobinska_PRA_86_063823_2012}
{M.~Stobi\'{n}ska, F.~T\"oppel, M.~Zukowski, M.~V.~Chekhova, G.~Leuchs, and N.~Gisin,}
\newblock Phys. Rev. A {\bf 86}, 063823 (2012).

\bibitem{Masalov_Klyshko8}
A.V.Masalov,
\newblock {\it Fine structure of quasiprobability function in quantum polarization
  tomography},
\newblock 8th D.N.Klyshko workshop, 2013.

\bibitem{Braunstein2005}
{S.L.Braunstein, P.van Loock},
\newblock Rev. Mod. Phys. {\bf 77}, 513 (2005).

\end{thebibliography}

\end{document}